\documentclass[oldversion]{aa}

\usepackage{epsfig}
\usepackage{graphics}
\usepackage{float}
\usepackage{amsmath}
\usepackage{multirow}
\usepackage{longtable}
\usepackage{rotate}
\usepackage{array}
\usepackage{subfigure}
\def\kms{\ifmmode{\rm km\,s^{-1}}\else\hbox{$\rm km\,s^{-1}$}\fi}

\setlongtables

\begin{document}

\title{Mass fluxes for O-type supergiants with metallicity $Z = Z_{\sun}/5$.}

\author{L.B.Lucy}
\offprints{L.B.Lucy}
\institute{Astrophysics Group, Blackett Laboratory, Imperial College 
London, Prince Consort Road, London SW7 2AZ, UK}
\date{Received ; Accepted }

\abstract{A code used previously to predict O-star mass fluxes as a function of
metallicity is 
used to compute a grid of models with the metallicity of the 
Small Magellanic Cloud (SMC). These models allow
mass-loss rates to be derived by interpolation for all
O-type supergiants in the SMC, with the possible exception of
extremely massive stars close to the Eddington limit.
}

\authorrunning{Lucy}
\titlerunning{SMC  mass-loss rates}
\maketitle

\section{Introduction}

A recent paper by Bouret et al. (2015) is a major contribution to our
understanding of mass loss by O-type supergiants at
low metallicity ($Z$). 

Earlier papers by Tramper et al. (2011,2014) analysing optical spectra
of extragalactic O stars in low-$Z$ environments 
derive mass-loss rates which, they claim, challenge the current paradigm 
of massive-star
evolution, both in the local universe as well as at cosmic distances. 
They make this claim because their estimates exceed predicted
rates for radiatively-driven winds. The implication, therefore,
is that a new, unknown mass-loss mechanism operates, and that
evolutionary tracks for massive stars require revision.

However, the Bouret et al. analyses of the far-UV HST/COS spectra of 
three of these stars decisively 
contradicts - see their Fig.5 - the rates found 
by Tramper et al.  
Moreover, the origin of the discrepacy is fully understood:
Tramper et al had perforce to rely on weak optical signatures of mass loss, 
namely the partial infilling of the H $\alpha$ and 
He {\sc ii} 4686 \AA\ absorption lines as a result of 
recombinations in the winds. But these diagnostics
have long been known (e.g., Lucy 1975) to overestimate mass-loss rates
because of wind clumping, an
effect neglected by Tramper et al.

With the revised 
rates of Bouret et al., there is now no case for a mass-loss mechanism
other than radiative driving, especially since their 
estimates are in good agreement with previously-computed mass fluxes 
(Lucy 2012; L12). This is an important conclusion since the Tramper et al. 
rates could imply a $Z$-independent mass- and angular momentum loss mechanism
that might even operate for Population III stars.

Nevertheless, Bouret et al. emphasize that this conclusion is based on
just three stars,
so that confirmation from a larger sample is desirable. To this end,
they promise a subsequent paper analysing archived spectra of O stars
in the Small Magellanic Cloud (SMC). To support this effort, 
this paper reports mass fluxes for models with SMC metallicity,
$Z = Z_{\sun}/5$.

Throughout this paper, mass flux in units   
gm s$^{-1}$ cm$^{-2}$  is denoted by $J$, whence the mass-loss rate
$\Phi = 4 \pi R^{2} J$ in solar masses per yr is given by
\begin{equation}
  \log \: \Phi  = \log \: J + 2 \: \log \: R/R_{\sun} - 3.015 
\end{equation}

\section{Computing mass loss rates}

In the light of the Bouret at al. (2015) analysis, this section 
briefly comments on two methods of predicting $\Phi$.  

\subsection{The Monte Carlo method}

In addition to the disagreement with Tramper et al., the $\Phi$'s
determined by Bouret et al. are $\sim 1$ dex below - see their Fig. 9 - those 
derived from
the Vink et al. (2001) fitting formula. Given that similarly large
overpredictions occur in the {\em weak-wind} domain (Marcolino at al. 2009,
Lucy 2010a; L10a) the Monte Carlo (MC) method
used by Vink et al. would appear to be discredited. 
This merits discussion.

The semi-empirical MC method used by Vink et al. was introduced by 
Abbott \& Lucy (1985; AL85) to 
investigate the differential effect of multi-line scattering, with the aim
of explaining the high $\Phi$'s  of Wolf-Rayet stars 
(Lucy \& Abbott 1993). The method is well-founded physically and 
should give reliable $\Phi$'s provided that, in the supersonic wind,
1) radiative driving is the 
acceleration mechanism, and 2) that the interaction of radiation and matter
is accurately modelled. This second point is where difficulties arise.

Detailed modelling of FUV line profiles - see Sect.8.2 of Bouret et al.
(2015) and references therein - demonstrates that the standard model 
of a homogeneous wind with a laminar outflow obeying a monotonic velocity law 
is contradicted: 
parameterized descriptions of 
severe clumping and highly
supersonic turbulence must be incorporated to fit the profiles.
Since these effects impact on the transfer of momentum from radiation
to matter, they evidently must be included in the MC method. Accordingly,
when using this method, the predicted emergent spectrum - see Fig.2 in
AL85 - should be compared to an observed spectrum to see if the chosen
clumping and turbulence parameters reproduce the observed P Cygni line 
profiles. 
\subsection{Prediction}
In principle, the structure of radiatively-driven winds can
be predicted from the equations of radiation gas dynamics. The results
could then be used to eliminate parameterized phenomenological
models. But this requires combining a treatment of 3-D, time-dependent, 
multi-frequency,  
non-LTE radiative transfer with that of 3-D,
time-dependent gas dynamics with 
multiple shocks. This is way beyond our capabilities, and so
quantitative prediction from first principles is not possible. 
   
But note that the supersonic wind arises from a stable stellar envelope.
Accordingly, as the Mach number $m \rightarrow 0$, the wind solution converges
to that of a 1-D, static photosphere in mechanical, statistical and thermal
equilibrium - i.e, to a structure that is eminently computable.

Plausibly, the transition from computable to non-computable occurs
early in the supersonic zone with the growth of instability
(Lucy \& Solomon 1970, LS70; Owocki et al. 1985), in which case the sonic point 
lies within the computable domain.   
A first-principles treatment of transonic flow is then feasible, and this
includes a determination of the mass flux $J$ as an eigenvalue.  
This assumption of computability is the fundamental basis of the
moving reversing layer (MRL) method
used to compute the low-$Z$ mass fluxes (L12) tested by Bouret et al. (2015) 

However, if extreme clumpiness and multiple shocks already arise when the
bulk motion is still subsonic, then the MRL method is undermined, and this 
should evidence itself in predictions that conflict with observations.

\subsection{The MRL method}

The MRL method is an updating of the crude treatment of dynamical reversing
layers given in LS70. The model was initially updated 
(Lucy 2007; L07) in order to investigate claims 
(Bouret et al. 2005; Fullerton et al. 2006)
that the theory of radiatively-driven winds overpredicts $\Phi$'s.
The same code was later used (L10a) to investigate the 
{\em weak-wind problem}.
Subsequently, the model was further refined (Lucy 2010b; L10b) by introducing
a flexible, non-parametric representation for the variation of 
the radiative  
acceleration due to lines $g_{\ell}$ with flow velocity $v$.  

The physical picture (LS70) motivating this model is that of the radiative 
expulsion of a stellar atmosphere's highest layers followed by a 
relatively gentle up-welling of deeper layers in response to the unbalanced
pressure. The MRL method assumes that this up-welling adjusts to a 1-D, 
time-independent outflow, with a smooth transition from sub- to supersonic 
velocity. This regularity constraint at the sonic point can only be satisfied
for a particular value of the mass flux $J$, which is therefore
an eigenvalue. 

As originally formulated, MRL models could be computed from first principles.
But there is ample evidence that the quasi-static reversing layers of 
O-stars are turbulent, and this is a phenomenon that we cannot yet
predict from first principles. Accordingly, the MRL models computed
here and previously follow Lanz \& Hubeny (2003) in assuming a canonical 
microturbulent velocity $v_{t} = 10 \:$ km s$^{-1}$ when computing
the Doppler widths of line profiles. In this regard, therefore, the
MRL models also acquire a semi-empirical aspect. This is a presently-
unavoidable departure from the ideal of calculation from first principles.

\section{Numerical solutions}

Mass fluxes $J$ are now computed for O-type supergiants in the SMC.

\subsection{Input data}

The composition has $N_{He}/N_{H} = 0.1$ and metals are reduced from solar
(Grevess \& Sauval 1998) by a factor 5.  
The included ions are as in Table 1 of Lanz \&
Hubeny (2003).  

The basic line data is from Kurucz \& Bell (1995). This is reduced to a 
working line list of the $\sim 10^{5}$ transitions relevant for the ions and 
atomic levels included in the TLUSTY models of Lanz \&
Hubeny (2003).

\subsection{Parameters}

The parameters 
are plotted on the $(\log T_{\rm eff}, \log g)$-plane in Fig.1, together
with evolutionary tracks for masses $25$ and $60 {\cal M}_{\sun}$
(Brott et al. 2011). The models thus 
encompass the $H$-burning phase of SMC O-stars with  
$ 25 \la {\cal M}/ {\cal M}_{\sun} \la 80$.
  
The points selected in Fig.1 are dictated by the availability
of TLUSTY models. As explained in L07 and L10a,
continuum fluxes and photospheric departure coefficients are derived from
models downloaded from the TLUSTY website. In effect, each MRL model
is grafted onto the corresponding TLUSTY model. This is justified
because significant departures from the static TLUSTY models occur
only when $m^{2} \ga 0.1$, and this corresponds to small optical depths. 
In this paper, the
downloaded data refers to the $S$ series of TLUSTY models, for which 
$Z = Z_{\sun}/5$.

For technical reasons, the TLUSTY models do not more closely approach
the Eddington limit than shown in Fig.1. Stars in the resulting gap - 
i.e., with ${\cal M}/ {\cal M}_{\sun} \ga 80$ -  may indeed exist in the SMC, 
and estimates of their $\Phi$'s are of great interest. But MRL predictions  
for such stars would require individual treatment.
Note that extrapolation with a scaling law is not justified - see Fig.7
in L10b.  

The MRL models are conveniently identified with the TLUSTY
notation. Thus $St400g350$ is the $S$ series model with 
$T_{\rm eff} = 40,000$K and $\log g( {\rm cgs}) = 3.50$.

\begin{figure}
\vspace{8.2cm}
\includegraphics{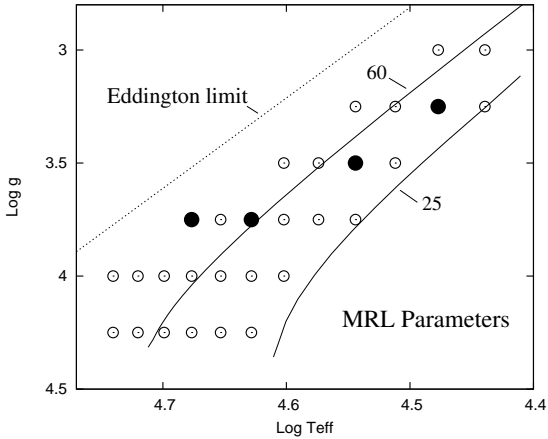}
\caption{Domain of interest in $(T_{\rm eff}, \log g)$-space. The filled circles
are loci of the O-star MRL models with $Z/Z_{\sun} \in (1/30,2)$ reported
in L12. The open circles are the loci of models with $Z/Z_{\sun} = 1/5$  
reported in this paper. The Eddington limit for static radiative envelopes
is shown as well as H-burning evolutionary tracks  (Brott et al. 2011) for 
masses $25$ and $60 {\cal M}_{\sun}$.}  

\end{figure}
\subsection{Mass fluxes}
The predicted $\log J$'s and other quantities of interest  
are given in Table 1 for each circled point 
{\bf $(T_{\rm eff}, \log g)$ } in Fig.1. 
For the filled circles, 
the values are the $Z/Z_{\sun}= 1/5$ models from L12, Table 1.

In addition to
$\log J$, Table 1 gives $\phi = c^{2} \Phi/ L$, the effective number of
strong lines; $\eta$, the percentage
of MC quanta that propagate through the MRL without interaction; and
$\zeta$, the percentage contribution of Fe and Ni to radiative driving by
lines in the MRL layer. These quantities are given as a function of $Z$
in L12, Table1 and    
are useful in understanding curve-of-growth effects and the $Z$-dependence of
the $J$'s - see
Sects. 2.3 and 3.3 in L12.

The new quantity in Table 1 is
\begin{equation}
  \Delta \: \log \: J =  \log \: J_{S} - \log \: J_{G} 
\end{equation}
where $\log \: J_{S}$ is the $S$ series value from Table 1 and
$\log \: J_{G}$ is the corresponding $G$ series ($Z/Z_{\sun} = 1$) value from 
L10b, Table 1.
The initial reason for tabulating this quantity is to identify possible
anomalies arising in the time-consuming, trial-and-error solution procedure.
But this quantity may also be useful for differential, model-insensitive 
comparisons of the spectra of Galactic and SMC O-type stars
- see Sect.3 in Bouret et al (2015).

For each value of $\log g$, the dependence of $\log J$ on $\log T_{\rm eff}$
is plotted in Fig.2. The corresponding plot for Galactic metallicity
$Z = Z_{\sun}$ is Fig.5 in L10b. As in that previous plot, Fig.2 shows the
expected trends that $J$ increases with increasing $T_{\rm eff}$ and
decreasing $g$.

\begin{figure}
\vspace{8.2cm}
\includegraphics{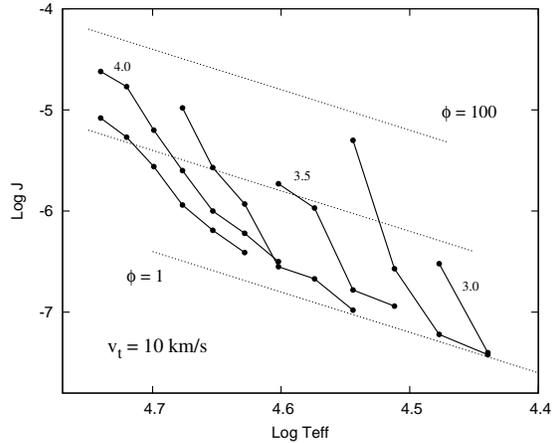}
\caption{Mass flux $J$ as a function of $T_{\rm eff}$ for 
$\log g = 3.00(0.25)4.25$. Lines of constant $\phi = c^{2} \Phi / L$
are also plotted. The metallicity $Z/Z_{\sun} = 1/5$,  and the microturbulent
velocity $v_{t} = 10 \:$km s$^{-1}$.}    

\end{figure}
\subsection{Failed solutions}
Inspection of Fig.1 or the $\Delta \log J$ values in Table 1
can lead one to suspect that some solutions are in error, perhaps reflecting
mistakes in the iterative, trial and error procedure adopted in the 
absence of an automatic search algorithm. Such suspected anomalies can be 
investigated with the procedure described in L12, Sect. 5.2. Specifically,
a value $J_{\dag}$ is selected that is more consistent with smooth variation,
and this value is fixed as the vector $g_{\ell}$ is adjusted.
If the best solution thus found fails to achieve consistency between the
input and output vectors $g_{\ell}$, then the solution search fails
and $J_{\dag}$ is contradicted. Typically, failure is evident by the
need to add or subtact momentum in the neighbourhood
of the sonic point - see L12, Fig.2.   

Failures of this kind are as follows: $St400g350$ at $J_{\dag} = -5.0$ dex;
$St400g375$ at $J_{\dag} = -6.3$ dex;
and $St375g350$ at $J_{\dag} = -6.25$ dex.

The implication of these failures is that these apparent anomalies  
indicate that the TLUSTY sampling in Fig.1 is somewhat too coarse to reveal 
all the real structure in the function $J(T_{\rm eff}, g)$.

\subsection{Particular cases}

Given the importance of the SMC as an accessible low-$Z$ environment,
it merits the attention of spectroscopists, as Bouret et al. (2015) 
recognize. This should then also prompt developers of stellar-wind codes to
compute models for O stars in the SMC.

Although comparisons with spectroscopic estimates are of primary 
importance, conflicts between different codes should also be identified and
understood. With respect to the MRL method, Table 1 provides data
for such comparisons. But in addition, Figs. 3 and 4 plot the converged
solutions of transonic flows for $St350g325$ and $St450g375$. In these
diagrams, the input assumption for the radiative acceleration due to lines 
(open circles) is compared to the estimates derived from the 
MC experiment (filled circles). Not only does output match input rather well, 
but the solutions exhibit smooth transitions from sub- to supersonic flow,
illustrating the constraints that determine the eigenvalues $J$.

\begin{figure}
\vspace{8.2cm}
\includegraphics{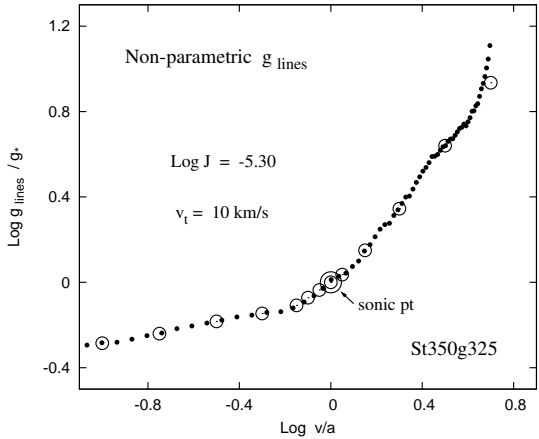}
\caption{MRL model St350g325. Radiative acceleration due to lines as a 
function of Mach number $v/a$ for
the transonic flow. The open circles define the assumed, non-parametric
input model; the filled circles are the MC estimates.}    

\end{figure}
\begin{figure}
\vspace{8.2cm}
\includegraphics{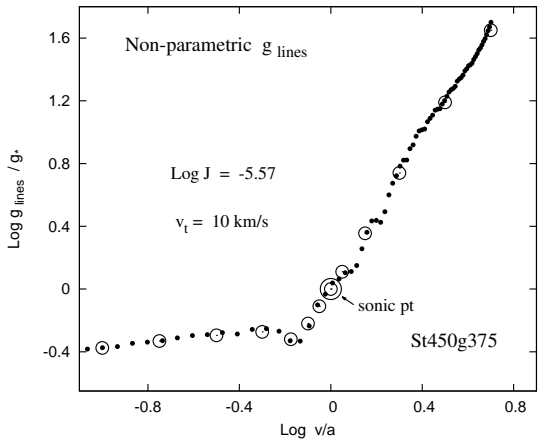}
\caption{MRL model St450g375.}   
\end{figure}
\subsection{Accuracy}
A lengthy discussion of the accuracy of the $J$'s is given in L10b, Sect.4.2
- see also L12, Sect.5.2.

Because of the steep gradients of $J(T_{\rm eff},g)$ - see Fig.2, 
a further source of error in predicting $J$  
for a particular star arises due to errors in the spectroscopic estimates
of $T_{\rm eff}$ and $g$. Such an error in the predicted $J$ is minimized if
$T_{\rm eff}$ and $g$ derive from diagnostics calibrated with
the predictions of TLUSTY atmospheres.

Yet another source of error is incompleteness of the line list. The working
line list (Sect. 3.1) comprises 87,469 transitions, of which 50\% have
$\log gf < -3.0$. Since incompletness will predominantly concern weak
transitions, it is instructive to exclude those with $\log gf < -3.0$ and
then to recompute $\log J$. This has been done for the models plotted in Figs.
3 and 4. For $St350g325$, the revised $\log J = -5.35$, so that including 
the weakest 50\% increases $J$ by 0.05 dex. For $St450g375$,
the corresponding increment in $J$ is again 0.05 dex. These calculations 
suggest that
incompleteness of the line list is less of a problem than other sources of 
error discussed above and in L10b. Nevertheless, future investigators of 
transonic flows should certainly update the line list.

\section{Conclusion}

With motivation provided by Bouret et al. (2015), the limited aim of this
paper has been to use an existing code (L10b, L12) to predict mass fluxes 
for O-type supergiants in the SMC. The results presented in Table 1
cover the full range of $T_{\rm eff}$ and $g$ expected for such stars.
Accordingly, for most observed stars, a prediction for its mass-loss rate
should be derivable by interpolation. Possible exceptions are
massive stars closer to the Eddington limit than the TLUSTY models allow
- see Fig.1. For such stars, prediction by extrapolation or by guessing
a scaling law is not recommended.

In comparing with spectroscopic estimates, a discrepancy for an individual
star might be due to errors in the assigned $T_{\rm eff}$ and $g$.
Of more significance would be a pattern of discrepancies that might point
the way to an improved theory.

Given the importance in understanding the $Z$-dependence of mass loss
by massive stars, this paper also includes data (Table 1) and 
diagrams (Figs. 3 and 4) for
comparisons with extant and future stellar-wind codes. Again, by analysing
discrepancies, we should eventually achieve greater 
predictive power.

\begin{table}

\caption{Computed mass fluxes $J$(gm s$^{-1}$ cm$^{-2}$) for 
$Z = Z_{\sun}/5$ and $v_{t} = 10 \:$ km  s$^{-1}$ .}

\label{table:1}

\centering

\begin{tabular}{c c c c c c c}

\hline\hline

  $T_{\rm eff}  (10^{3}$K)    &  $\log g$  & $\log J$ & $\Delta \log J$ & $\phi$ & $\eta (\%)$ & $\zeta(\%)$  \\

\hline
\hline

     55.0 &   4.25   & -5.08 & -0.60   & 14.4  &  76.6  & 38.8  \\

     52.5 &          & -5.27 & -0.53  & 11.2  &  75.8  & 33.6  \\

     50.0 &          & -5.56 & -0.28  &  7.0  &  74.5  & 25.7  \\

     47.5 &          & -5.94 & -0.42  &  3.6  &  72.5  & 18.7  \\

     45.0 &          & -6.19 & -0.35  &  2.5  &  71.2  & 14.9  \\

     42.5 &          & -6.41 & -0.46  &  1.9  &  71.5  & 14.2 \\

\cline{1-7}

     55.0 &   4.00   & -4.62: & -0.83  & 41.6    & 79.4  & 80.6  \\

     52.5 &          & -4.77  & -0.59 & 35.4    & 78.1  & 68.2  \\

     50.0 &          & -5.20  & -0.74 & 16.0    & 77.2  & 50.3  \\

     47.5 &          & -5.60  & -0.62 & 7.8    & 75.6  &  37.9 \\

     45.0 &          & -6.00  & -0.46 & 3.9    & 73.4  &  23.3 \\

     42.5 &          & -6.22  & -0.52  & 2.9     & 72.8  & 21.2  \\

     40.0 &          & -6.50  & -0.51 &  2.0   & 73.5  & 19.6  \\

\cline{1-7}

     47.5 &  3.75   & -4.98    & -0.67  &  32.6  & 58.6 & 69.9 \\   

     45.0 &         & -5.57    &  -0.71   & 10.4    & 76.8  & 56.2  \\

     42.5 &         & -5.93    & -0.79  & 5.7  & 74.5 & 38.7 \\

     40.0 &         &  -6.55   &  -0.83    & 1.8    & 76.2  & 20.8 \\

     37.5 &         &   -6.67   &  -0.63   & 1.7    & 76.1  & 19.7 \\

     35.0 &         &  -6.98    &  -0.12  &  1.1   & 74.8  &  7.8  \\

\cline{1-7}

     40.0 &   3.50  & -5.73  &  -1.31  & 11.5    & 75.3 &  57.1  \\

     37.5 &         & -5.97  &  -1.12   &  8.6     & 75.3  & 57.9  \\

     35.0 &         & -6.78  &  -0.68    &    1.8  & 76.4 & 23.2 \\

     32.5 &         &  -6.94  & -0.61   &  1.6    & 74.0 &  6.7   \\

\cline{1-7}

     35.0 &  3.25   &  -5.30 & -0.51  &  52.9   & 76.4 & 79.1   \\

     32.5 &         & -6.57:  & -1.10  & 3.8    & 77.1  & 33.9  \\

     30.0 &         & -7.22  &  -0.71   &    1.2  & 76.2 &  5.2 \\

     27.5 &         & -7.42: & -0.02   &  1.0   & 75.0  & 4.7  \\

\cline{1-7}

     30.0 &  3.00   & -6.52  & -1.12 &  5.9   & 76.5   &  48.8 \\
 
     27.5 &         &   -7.41: & -0.46  & 1.1    & 77.4 & 8.8  \\

\hline
\hline

\end{tabular}

\end{table}

\acknowledgement

I thank the referee, W.-R. Hamann, for a careful reading of the manuscript
and for thoughtful suggestions.

\end{document}